\documentstyle[12pt,titlepage,epsfig]{article}
\input epsf

\renewcommand{\theequation}{\thesection.\arabic{equation}}

\font\medio=cmr10 scaled \magstep2
\outer\def\beginsection#1\par{\medbreak\bigskip
      \message{#1}\leftline{\bf#1}\nobreak\medskip
\vskip-\parskip
      \noindent}

\def\laq{\raise 0.4ex\hbox{$<$}\kern -0.8em\lower 0.62
ex\hbox{$\sim$}}
\def\gaq{\raise 0.4ex\hbox{$>$}\kern -0.7em\lower 0.62
ex\hbox{$\sim$}}
\def\be{\begin{equation}}
\def\ee{\end{equation}}
\def\bees{\begin{eqnarray}}
\def\ees{\end{eqnarray}}

\def \pa {\partial}
\def \ra {\rightarrow}

\def \dH {\dot{H}}
\def \df {\dot{\phi}}
\def \ddf {\ddot{\phi}}
\def \fbar {\bar{\phi}}
\def \dfbar {\dot{\bar{\phi}}}
\begin{document}
\titlepage
\begin{flushright}
CERN-TH/97-228 \\
\end{flushright}
\vspace{7mm}
\begin{center}
{\bf MASSIVE STRING MODES AND NON-SINGULAR \\
         PRE-BIG-BANG COSMOLOGY}

\vspace{9mm}

Michele Maggiore\footnote{Permanent address:
{INFN and Dipartimento di Fisica, piazza Torricelli 2, I-56100 Pisa,
Italy.}} \\
{\sl Theory Division, CERN, CH-1211 Geneva 23, Switzerland} \\

\end{center}
\vspace{9mm}
\centerline{\medio  Abstract}

\vspace{2mm}
\noindent
Perturbative $\alpha '$ corrections to the low energy string effective
action have been
recently found to have a potentially regularizing effect on the
singularity of the lowest order pre-big-bang solutions. Whether they
actually regularize it, however, cannot be determined working at
any finite order in a perturbative expansion in powers of the string
constant $\alpha '$, because of scheme dependence ambiguities. 
Physically, these corrections are dominated by the
integration over the first few massive string states.
Very massive string modes, instead, can  have a regularizing effect
which is non-perturbative in $\alpha '$ and
which basically comes from the fact that in a gravitational field
with Hubble
constant $H$ they are produced with an effective Hawking temperature 
$T=H/(2\pi )$, and an infinite production rate would occur if
this temperature  exceeded the Hagedorn temperature.
We discuss technical and conceptual difficulties of this
non-perturbative regularization mechanism.

\vspace{5mm}

\vfill
\begin{flushleft}
CERN-TH/97-228 \\
August 1997\\
revised April 1998
\end{flushleft}

\newpage

\renewcommand{\theequation}{1.\arabic{equation}}
\setcounter{equation}{0}
\section {Introduction}
String theory is an appropriate framework for discussing the
singularities of
general relativity and in particular the big-bang singularity,
and a large number of works have been devoted to the study of
cosmological solutions of the low energy effective action of string
theory (see e.g.~[1-5]).
At lowest order in the string constant $\alpha '$,
the solutions of the equations of motion  still reach a singularity
at a finite value of time, say $t=0$, when we evolve the present state
of the Universe backward in time~\cite{GV}. This is not surprising
since, in the lowest order effective action, $\alpha '$ only enters as
an overall constant and therefore drops from the equations of
motion; therefore, there is no scale at which the singularity can be
regularized. For homogeneous fields, the low energy action
has a symmetry, scale factor duality, which relates different
solutions of the equations of motions~\cite{Ven,duality}. 
In the simplest case of an
isotropic Friedmann-Robertson-Walker (FRW) metric with scale factor
$a(t)$, and of
a vanishing antisymmetric tensor field $B_{\mu\nu}$, it reads
\be
a\ra\frac{1}{a},\hspace{5mm}\phi\ra\phi -2 d\log a\, ,
\ee
where $\phi$ is the dilaton field and $d$ is the number of spatial
dimensions. This symmetry can be generalized to a global $O(d,d)$
symmetry in the more general case of non-diagonal metrics and 
non-vanishing $B_{\mu\nu}$. Since it involves the dilaton field, it is a
string symmetry, with no counterpart in Einstein gravity. Combining
scale factor duality with time reversal, it is possible to associate
with every `post-big-bang' solution defined for $t>0$ a `pre-big-bang'
solution defined for $t<0$, through the transformation $a(t)\ra 1/a(-t)$.
Of course, at lowest order in $\alpha '$, the pre-big-bang solution
also runs into a singularity as $t\ra 0^-$. It is therefore natural to
ask whether the inclusion of corrections allows a smooth transition
between the pre-big-bang and the post-big-bang solutions, providing a
non-singular cosmological model. 

The effective action of string theory
has two different types of perturbative
corrections: higher orders in $\alpha '$, which are genuinely stringy
corrections related to the finite extent of the string, and loop
corrections, which carry higher powers of $e^{\phi}$. Perturbative
$\alpha '$ corrections provide a scale for the regularization of the
singularity and have been studied in~\cite{GMV}, where it has been
found that the equations of motion in
the case of constant curvature and linear dilaton reduce, at 
all orders in $\alpha '$, to a set of $(d+1)$ algebraic equations in
$(d+1)$ unknowns. If these algebraic equations have a real solution,
this can act as late time attractor of the lowest order pre-big-bang
solution, as  has been shown on specific examples at 
$O(\alpha ')$. The singularity is then replaced by a
phase of De~Sitter inflation with linearly growing dilaton.  
To complete the transition to the standard radiation-dominated {\em FRW}
model (the graceful exit problem)
$O(e^{\phi})$ corrections must play a crucial
role~\cite{gra1,gra2,BM,BM2}. 
 
In this paper we examine again the role of $\alpha '$ corrections in
string cosmology. In sect.~2, after recalling  the results of
ref.~\cite{GMV} and extending them to the case of compact dimensions
(sect.~2.1), 
we discuss in some detail the scheme dependence of the results
(sect.~2.2)  and the
relation with scale factor duality (sect.~2.3). Unfortunately, to
verify that the mechanism proposed in~\cite{GMV} does take place would
require the knowledge of a beta function at all orders in $\alpha '$. Therefore,
in sect.~3 we try to obtain a better understanding of the physics
behind the perturbative corrections and of the general mechanism of
smoothing of singularities in string theory. We will relate the
perturbative $\alpha '$ corrections
to the integration over the first few excited
string states (sect.~3.1); instead, the mechanism of singularity regularization
is related to the exponential growth of the density of states; it is
therefore  basically due to very massive string modes.

In sect.~4 we  study the effect on the lowest order solution
of the backreaction due to very massive string modes produced by the
gravitational field. For technical reasons, that we will discuss in
sect.~4, it turns out to be quite difficult to present an explicit
non-singular solution. The result also suggests that treating massive modes
production as a backreaction, in a full stringy regime, is not adequate.
However, from the equations that include the
creation of massive modes, we will be able to obtain at least a
qualitative understanding of the role of massive modes production
and of their possible regularizing effect.
In sect.~5 we present the conclusions and we discuss possible
phenomenological implications of a string phase with highly excited
massive modes.

\renewcommand{\theequation}{2.\arabic{equation}}
\setcounter{equation}{0}

\section { Perturbative $\alpha '$ corrections}

\subsection{The constant curvature solution}

Let us  recall the results of ref.~\cite{GMV}.
Including  first order corrections
in $\alpha '$, a possible form of the effective action, in the string
frame,  is~\cite{azione,FT}
\be\label{act0}
S=-\frac{1}{2\lambda_s^{d-1}}\int d^{d+1}x\sqrt{-g}\, e^{-\phi}
\left[ R+(\nabla \phi )^2-\frac{k\alpha '}{4}R_{\mu\nu\rho\sigma}^2
\right]\, ,
\ee
where $\lambda_s$ is the string length, $\lambda_s\sim\sqrt{\alpha '}$,
$k=1,1/2$ for the bosonic
and heterotic string, respectively, and we have neglected
the antisymmetric tensor field. For type II strings $k=0$, and the
first correction starts at order $R_{\mu\nu\rho\sigma}^4~$\cite{GW}.
Performing the field redefinition $g_{\mu\nu}\ra g_{\mu\nu}+
(k\alpha ')\delta g_{\mu\nu}$, $\phi\ra\phi +(k\alpha ')\delta\phi$ with
\be\label{redef1}
\delta g_{\mu\nu}= R_{\mu\nu}
-\partial_{\mu}\phi\pa_{\nu}\phi+g_{\mu\nu} (\nabla\phi )^2,
\hspace{4mm}
\delta\phi =\frac{1}{4}\left[ R+(2 d-3)(\nabla\phi )^2\right]
\ee
and truncating at first order in $\alpha '$ gives the
action
\be\label{act}
S=-\frac{1}{2\lambda_s^{d-1}}\int d^{d+1}x\sqrt{-g}\, e^{-\phi}
\left[ R+(\nabla \phi )^2-\frac{k\alpha '}{4}\left( R_{GB}^2-
(\nabla\phi )^4\right)
\right]\, ,
\ee
where $R_{GB}^2=R_{\mu\nu\rho\sigma}^2-4R_{\mu\nu}^2+R^2$ 
is the Gauss-Bonnet term. This form is
particularly convenient because higher order derivatives in the
$\alpha '$ correction cancel after integrations by part~\cite{Zwi}.
We will discuss in some detail field redefinitions in sect.~2.2.

We now restrict ourselves
to a homogeneous and isotropic {\em FRW} background,
$ds^2=N^2(t)dt^2-a^2(t)d{\bf x}^2$ and homogeneous dilaton $\phi
=\phi (t)$. Varying the action with respect to $a,\phi$, we get two
dynamical equations of motion. Introducing $H=\dot{a}/a$ 
and specializing to $d=3$ for notational simplicity, they read (in
units $k\alpha '=1$):
\be\label{din1}
-6 \dH (1+H^2)+2\ddf \left( 1+\frac{3}{2}\df^2\right) -12
H^2+6H\df-\df^2-\frac{3}{4}\df^4-6H^4 +3 H\df^3=0
\ee
\be\label{din2}
-12\dH (1-H\df )+6 \ddf (1+H^2)-18 H^2+12 H\df
-3\df^2+\frac{3}{4}\df^4 +12H^3\df-6H^2\df^2=0\, .
\ee
The variation with respect to the lapse
function $N$ gives, instead, a constraint on the initial values: 
\be\label{vin}
6H^2-6H\df+\df^2-6H^3\df+\frac{3}{4}\df^4=0\, .
\ee
The constraint is conserved by the dynamical equations of motion, and
is therefore satisfied at any time if it is satisfied at the initial
time. We can now try to solve eqs.~(\ref{din1}-\ref{vin}) with the
ansatz $H=$ const $ =y$, $\df =$ const $=x$. The ansatz reduces the
three differential
equations to algebraic equations in $x,y$. At first sight we
have three independent equations for two unknown variables.  
However, reparametrization invariance gives one relation between these
equations such that if the constraint equation and, say, the
equation obtained with a variation with respect to $a$
(or, in the general anisotropic case, the
$d$ equations obtained with a variation  with respect to $a_i$,
$i=1,\ldots\, , d$) are
satisfied, then the equation obtained with a variation with respect to
$\phi$ is automatically satisfied.

In the generic anisotropic case with scale factors $a_i$ $(i=1,\ldots
\, ,d)$,  the ansatz therefore reduces the system of $(d+2)$ differential
equations to $(d+1)$ algebraic equations  in $(d+1)$
variables $H_i=y_i,\df =x$, and this is true at all orders in $\alpha '$.

These algebraic equations are nothing but the requirement that there is a
zero in the beta functionals of the underlying sigma model, when the
background is specialized to the form of our ansatz,
and we will therefore write them as $\beta_i({\bf g})=0$, where
$i=1,\ldots ,(d+1)$ and ${\bf g}=(x,y_1,\ldots ,y_d)$.

It is easy to generalize this result to the case of compact
dimensions. Let us first consider  the case in which the compact space
is spatially curved, e.g. consider four-dimensional  space $M^4$ times 
$S^2$, with metric
\be
ds^2=N^2(t)dt^2-\sum_{i=1}^3a_i^2(t)dx_i^2-b^2(t)\,
(d\theta^2+\sin^2\theta d\phi^2)\, .
\ee
(The extension to more general cases will be obvious.)
Writing down the equations of motion we see that now $b$ 
enters not only in the combination $\dot{b}/b$, but also 
through terms $\sim1/b^2$, which
are due to the curvature of the
sphere. Therefore the ansatz $\dot{a_i}/a_i=$ constant, $\df =$ constant
can only be consistent if $b=$ const, or $\dot{b}/b=0$. 
Again for the ansatz $\dot{a_i}/a_i =y_i, 
\df=x, b={\rm const}$
the relation between equations of motion derived from
reparametrization invariance eliminates one equation and we have a
number of algebraic equations equal to the number of 
variables $y_i,x,b$.

The same happens if,  instead, we compactify on a torus. In this case the
explicit dependence on the scale factor, which forces it to be
constant, comes from the winding modes, whose energy grows
with the scale factor.

In conclusion, the existence of a solution with  $H_i,\df$ constant
(and $b$ constant, for compact dimensions)
depends on whether the  algebraic equations discussed above have real
solutions. For the action~(\ref{act}) this is indeed the case, and the
solution turns out to be an attractor of the lowest order pre-big-bang
solution, which is therefore regularized~\cite{GMV}. However, the
inclusion of higher orders in $\alpha '$ or field redefinitions of the
type used in eq.~(\ref{redef1}) produce different algebraic equations,
which may or may not have real solutions. These issues, which were
already noted in ref.~\cite{GMV}, will be further discussed in the next
section.

\subsection{Scheme dependence of the results}
In principle, we would like to know the beta functions 
$\beta_i({\bf g})$ exactly, while what 
we have is a perturbative expansion  in powers of $\alpha '$. 
Somewhat
optimistically, one might still hope to find a zero which, in
units $k\alpha '=1$, is located at $g_i\ll 1$, thus justifying a
perturbative treatment. Unfortunately, an even more fundamental
obstacle stands in the way. The problem is that, if we work at finite
order in $\alpha '$, the coefficients of the algebraic equation, or of
the expansion of the functions $\beta_i$, are subject to 
ambiguities. 
A straightforward way to understand this point
is to observe that we can perform 
fields redefinitions that mix different orders in $\alpha '$. 
The most general form of such redefinitions, at
order $\alpha '$, is~\cite{MT}
$g_{\mu\nu}\ra g_{\mu\nu}+(k\alpha ')
\delta g_{\mu\nu}$, $\phi\ra\phi +(k\alpha ')\delta\phi$, with
\be\label{redef2}
\delta g_{\mu\nu}= a_1 R_{\mu\nu}
+a_2\partial_{\mu}\phi\pa_{\nu}\phi+a_3g_{\mu\nu} (\nabla\phi )^2
+a_4g_{\mu\nu}R+a_5g_{\mu\nu}\Box\phi\, ,
\ee
\be
\delta\phi = b_1R+b_2(\nabla\phi )^2+b_3\Box\phi\, .
\ee
It is not necessary to include a term $\nabla_{\mu}\pa_{\nu}\phi$ in 
eq.~(\ref{redef2}) because it can be reabsorbed by a general
coordinate transformation~\cite{MT}. After this redefinition, the new
action, truncated at order $\alpha '$, is (in units $k\alpha '=1$)
$$
S=-\int d^{d+1}x\sqrt{-g}\, e^{-\phi}
\left[ R+(\nabla \phi )^2-\frac{1}{4}R_{\mu\nu\rho\sigma}^2
+c_1R_{\mu\nu}R^{\mu\nu}+c_2R^2+c_3(\nabla\phi )^4+\right.
$$
\be\label{act2}
\left. +c_4R^{\mu\nu}
\pa_{\mu}\phi\pa_{\nu}\phi+c_5R(\nabla\phi )^2+c_6R\Box\phi
+c_7\Box\phi(\nabla\phi )^2+c_8(\Box\phi )^2
\right]\, ,
\ee
with the coefficients $c_i$ functions of $a_i,b_i$. (We have
eliminated terms that can be reduced to the above terms after
integrations by part or use of Bianchi identity). Despite the fact
that we have eight coefficients $c_i$ and eight parameters $a_1,\ldots
\, ,a_5, b_1,b_2,b_3$, we cannot fix the $c_i$ at arbitrary values since,
from the explicit expression of the $c_i$ as functions of $a_1,\ldots
\, ,b_3$, one finds that they satisfy a relation
$c_2+c_3+c_7+c_8=c_5+c_6$. Within this constraint, however, the $c_i$
can be chosen at will with the appropriate field redefinitions. The
existence of a relation between the $c_i$
means that there is a one-parameter family of field redefinitions
which, at order $\alpha '$, leaves the action invariant. It is readily
found to be
\be
\delta g_{\mu\nu}=\zeta g_{\mu\nu}\left( R-(\nabla\phi )^2+2\Box\phi
\right)\, ,
\ee
\be
\delta\phi =\zeta\left( \frac{d-1}{2}R-\frac{d+1}{2}(\nabla\phi )^2
+d\Box\phi \right)\, ,
\ee
with $\zeta$ a real parameter.
We now ask whether the existence of a zero of the functions
$\beta_i$, at a given order in $\alpha '$, is affected by the field
redefinitions, and the answer is positive. For instance, we can fix $c_1=1,
c_2=-1/4,c_3=1/4$, and vary $c_4$ (which does not enter the relation
between the $c_i$)  setting all others $c_i$ to zero.
For $c_4=0$ we have the action used in~\cite{GMV},
and there is a zero of the beta functions. In $d=3$, for the isotropic
case, it is located at $x\simeq 1.404,y\simeq 0.616$.
Increasing $c_4$ this zero disappears
(escaping at infinity in the $(x,y)$ plane) at a critical value 
$c_4\simeq 0.05$.

Therefore, while we expect that, at all orders in $\alpha '$, the
existence of a zero in the beta functions is independent of  field
redefinitions, this is not true
for the truncation at any finite order in $\alpha '$.

There are other useful ways to understand the 
existence of ambiguities in the
perturbative expansion of the
functions $\beta_i$, and these different points of view
are believed to be equivalent.
First, from the point of view of the underlying sigma
model, they are due to the dependence of the perturbative coefficients
of the beta functionals on the
renormalization scheme. This dependence 
 starts at two loops, i.e. from the terms
$\sim R^2$ in the action, eq.~(\ref{act2}).

The effective low energy action, on the other hand, is constructed in
such a way as to reproduce the string theory $S$-matrix elements. 
From this point of view, the ambiguities in the coefficients come
from the fact that some coefficients cancel in the computation of  
on-shell amplitudes~\cite{DR,MT,FOTW}. Suppose for instance that we want to
fix the coefficients of the operators
$R_{\mu\nu\rho\sigma}^2,R_{\mu\nu}^2$ and $R^2$ in the effective
action. We would then expand these operators
 around the flat metric, $g_{\mu\nu}=
\eta_{\mu\nu}+h_{\mu\nu}$ and compute their contribution to the
3-point and 4-point amplitudes. However, the $\sim h^3$ part of
$\sqrt{-g}R_{\mu\nu}R^{\mu\nu}$ and of $\sqrt{-g}R^2$ vanish;
therefore, the coefficients $c_1,c_2$ in 
the action~(\ref{act2}) cannot be determined from the
knowledge of the  string amplitude with three on-shell gravitons. 
Both $\sqrt{-g}R_{\mu\nu}R^{\mu\nu}$ and  $\sqrt{-g}R^2$ 
are non-vanishing when expanded at order  $h^4$. However, the string
amplitude includes one-particle reducible graphs with graviton
exchange;  to reproduce it we must therefore sum
the contact and the exchange terms derived from the effective
action. In the sum, the coefficients
$c_1,c_2$ cancel (see~\cite{FOTW} for a detailed computation), and
so they cannot be determined from the comparison with string
amplitudes. The coefficient of $R_{\mu\nu\rho\sigma}^2$ is instead fixed by this
procedure, at the value $-1/4$. 

\subsection{Relation with scale factor duality}
Scheme dependence should not change physical results if we had the exact
expression for the beta functions. It enters into play when we
truncate at finite order in $\alpha '$. When  scheme dependence
appears,  some scheme will be `better' than others, in the sense
that the results obtained at finite order in this scheme
will be closer to the exact result.  Since scale factor duality plays
an important role in motivating the cosmological model that we are
discussing, one might hope that a scheme that respects scale factor
duality at a given order in $\alpha '$ will be better, in the above
sense.

Scale factor duality has been generalized to $O(\alpha ')$ 
in refs.~\cite{Mei,MK}. 
The analysis of ref.~\cite{Mei} is very general, and
refers to the full $O(d,d)$ symmetry.  The result of
ref.~\cite{Mei} is that there is one, and only one action invariant
under scale factor duality at order $\alpha '$, and it is given by the
action~(\ref{act2}) with $c_1=1,c_2=-1/4,
c_3=1/4,c_4=1, c_5=-1/2,c_6=0,c_7=-1/2,c_8=0$. At
the same time, the duality transformation must acquire $\alpha '$
corrections. 
If we restrict to {\em FRW}
metrics and vanishing antisymmetric tensor field, then 
$a_i\ra 1/a_i$ (or $\log a_i\ra-\log a_i$) must be generalized as
\be\label{dua}
\log a_i\ra -(\log a_i)-k\alpha '
\left(\frac{\dot{a}_i}{a_i}\right)^2\, .
\ee
The transformation of the dilaton is fixed from the condition that 
$\phi -\sum_i\log a_i$ is invariant.
In $d=3$, for isotropic metric, the
 equations  $\beta_i(x,y)=0$ for this dual action have  two real
solutions\footnote{Real solutions for this action
exist for any $d$. This can be proved numerically up to large values of $d$,
where the results matches with a large $d$ expansion.}
 at $(x\simeq 1.526,y\simeq 1.913)$ and at 
$(x\simeq 6.201,y\simeq 1.931)$ (plus the solutions $(-x,-y)$, which
are always associated with the solution $(x,y)$).
The existence of a pair of solutions (plus the sign-reversed pair)
is related to the scale factor duality of the action. However, 
for the action that we are considering,  scale factor duality is valid
only at order $\alpha '$, and not exactly, since the transformation
(\ref{dua}), when applied to the $O(\alpha ')$ terms of the action, generates 
$O(\alpha '^2)$ terms that are truncated. Thus the two solutions
are not exactly dual to each other.

Integrating the equations of motion numerically, one finds~\cite{GMV}
 that these solutions do not act as a late time attractor of the
lowest order solution, which still diverges at some finite
value of time. 

To further explore the relation with duality
we have tried  a different generalization
of duality at $O(\alpha ')$. In fact, the action found in ~\cite{Mei}
and the form of the duality transformation~(\ref{dua}) are uniquely
fixed only if we consider the general case of non-isotropic metrics. If
however we restrict to isotropic metrics, a transformation
\be\label{dua2}
\log a\ra -(\log a)-\lambda k\alpha '
\left(\frac{\dot{a}}{a}\right)^2\, ,
\ee
with arbitrary $\lambda$, leaves the action (\ref{act2}) 
invariant at order  $\alpha '$ if  the coefficients $c_i$ are
$c_1=1,c_2=-1/4,c_3=1/4+(\lambda -1)/18,
c_4=1+(\lambda -1)/9, c_5=-c_4/2,c_6=0,c_7=-2c_3,c_8=0$.
(The relation between the $c_i$ is still satisfied.)
For $\lambda =1$ we recover the action
and the transformation  of ref.~\cite{Mei}. However, 
for every transformation~(\ref{dua2}) with given $\lambda$ we now
have one action that is
invariant.\footnote{Basically, we have a much less rigid structure
because many expressions which are independent in the anisotropic case, 
such as $\sum_iH_i^2$ and $(\sum_i H_i)^2$, collapse to the same expression in
the isotropic case, and we can arrange cancellations between them.} 
In a sense, this enlarged family of transformations is
less interesting than the transformation found in~\cite{Mei}, because
the latter is the only one that can be generalized to the anisotropic
case and to the inclusion of the antisymmetric tensor field. However,
this family of transformations also includes the case $\lambda=0$, 
i.e. the transformation $a\ra 1/a$,
without $\alpha '$ corrections, which is appealing for its
simplicity. Furthermore, since this transformation does not generate
higher orders in $\alpha '$, it is an exact invariance of the action
(\ref{act2}), with the appropriate choice of $c_i$, rather than an
invariance at $O(\alpha ')$.

We have therefore studied the equations of motion for the action
with $\lambda =0$ duality. We have found that
the functions $\beta_i(x,y)$ have a pair of zeros (plus the sign
reversed pair), and these zeros
are related by exact duality invariance, as it should. But, again,
these solutions
do not act as late time attractors of the lowest order  solution,
which instead runs into a singularity. The type of singularity is however
different from the one encountered for the case $\lambda =1$.
It is interesting to understand in some detail
what happens. The dynamical equations
of motion can be written in matrix form
\be
A \pmatrix{  \ddot{\phi}\cr  \dot{H}\cr}
= \pmatrix{  f_1\cr  f_2\cr}
\ee
where $A$ is a $2\times 2$ matrix that depends on $H,\df$, and $f_1,f_2$
are functions of $H,\df$. At a finite value of time we find that $\det
A=0$, and therefore the equations become singular. 
As a simpler example of a similar situation, consider the differential
equation $(1-h)\dot{h}=h$ for some function $h(t)$. For small $h$ the
solution grows like $e^t$, until the 'determinant' $(1-h)$ vanishes,
so the situation is similar to our case. Here however we can integrate
the equation exactly, and find $t(h)=\log h -h$. Above a critical
value  $t=-1$, this relation cannot be inverted while,
 below this value, there are
two branches $h_+(t)$ and $h_-(t)$ that merge at $t=-1$.

This suggests that what happens in our case is that the doubling of
solutions due to duality transformations with $\lambda =0$ produces
pairs of solutions of
the dynamical equations of motion, which
merge at some critical value
of time, and thereafter move into the complex plane. (These solutions
are not dual to each other. One corresponds to a small initial value of $H$
and $\dot{H}>0$, and the other to $H$ large, and $\dot{H}<0$,
while duality changes $H\ra -H$). If this conjecture is correct,
then it appears that duality works against the regularization
mechanism. In any case, neither a scheme that respects
duality with $\lambda =1$ nor a scheme that respects duality with $\lambda
=0$ provides a realization of the regularization mechanism, 
at $O(\alpha ')$.

\renewcommand{\theequation}{3.\arabic{equation}}
\setcounter{equation}{0}

\section{The physics behind  the regularization of the singularity}
The conclusion of the previous sections is that perturbative 
$\alpha '$ corrections can in principle regularize the singularity; 
however, to determine whether this actually happens we
should know the functions $\beta_i$ at all orders in $\alpha '$,
because results at finite order are scheme dependent. The attempt to
fix the scheme using scale factor duality as a guiding principle does
not give encouraging results. We  therefore ask whether there
are general physical principles that suggest that the singularity is
indeed regularized. Generally speaking, we expect that
string theory eliminates all unwanted singularities. However, is the
physical mechanism that eliminates the singularity related to
perturbative $\alpha '$ corrections? Or should we look for a different
type of corrections? In this section we   discuss this
question. 

\subsection{Effective action and massive string modes}
As it is well known, the low energy action of string theory can be
obtained either by computing the beta functions of the sigma model or
from the requirement of reproducing the $S$-matrix elements. The two
methods are equivalent, as was proved in~\cite{BNY}. Still,
the physical pictures underlying the two computations are quite
different. In particular, in the first method the massive string modes
seem to play no role at all. 

In fact,  what one does in order
to include massive modes at the sigma model
level  is the following. The sigma model
action is written including the interaction with the background fields
representing the massive modes~\cite{FT,Tse},
\bees
S&=\frac{1}{\alpha '}
\int d^2\sigma\sqrt{-h}&\left[ g_{\mu\nu}(X)h^{ij}\pa_iX^{\mu}\pa_jX^{\nu}
+B_{\mu\nu}(X)\epsilon^{ij}\pa_iX^{\mu}\pa_jX^{\nu}+
\phi (X) ^{(2)}R+\right.\nonumber\\
& &\left. \hspace*{3mm}+
\frac{1}{\alpha '}F_{\mu\nu\rho\sigma}(X)h^{ij}h^{kl}
\pa_iX^{\mu}\pa_jX^{\nu}\pa_kX^{\rho}\pa_lX^{\sigma}
+\ldots\right]\, ,
\ees
where together with the metric, dilaton and antisymmetric field, we
have displayed an example of a field at the next mass level.
Once we include massive fields at the first excited level, the whole
tower of massive fields must be included, since they are associated
with non-renormalizable interactions from the point of view of the
2-dimensional theory, and all possible terms are in principle
generated by the sigma model loop (i.e. $\alpha '$) expansion. 

However, when we compute higher loops in the beta
functionals of the  massless modes,
the  massive modes play no role: the operators
associated with massive modes are non-renormalizable and do not mix
with the massless sector. More generally, to renormalize background
fields at a given mass level, we need only fields at the same or lower
mass level~\cite{BFLP}.

However, since in string theory we have an infinite tower of massive
modes, while the effective low energy action is written in terms of
massless fields only, it is clear that we have somehow integrated over
the massive modes. This physical interpretation is confirmed by
the computation of the effective action through the comparison with
$S$-matrix elements.
In this case the effective action, at all orders in
$\alpha '$ but without $e^{\phi}$ corrections, is obtained by
requiring that it reproduces the
string amplitudes at genus zero, with an arbitrary number of
insertions of vertex operators for the massless fields. In the field
theory language, we expect that the $\alpha '$
corrections to the  amplitudes can be represented as a sum of 
tree level exchange graphs with  massive string modes
in the intermediate state. This is
indeed the case, as  can be read from  the computation of Gross and
Witten~\cite{GW} of  graviton scattering in type II superstring
theory. In this case the tree amplitude is~\cite{GS,GW}
\be\label{ampl}
A=\frac{G_N^2}{128}\left[ 
\frac{\Gamma (-s/8)\Gamma (-t/8)\Gamma (-u/8)}
{\Gamma (1+\frac{1}{8}s)\Gamma (1+\frac{1}{8}t)\Gamma (1+\frac{1}{8}u)}
\right] K(\epsilon^{(i)},k^{(j)})\tilde{K}(\epsilon^{(i)},k^{(j)})\, ,
\ee
where $G_N$ is the gravitational constant, $s,t,u$ the Mandelstam
variables and $K,\tilde{K}$ are kinematical factors.

Expanding the term in brackets for small momenta, 
one gets a leading term  $-2^9/(stu)$, which, when inserted back 
in eq.~(\ref{ampl}), gives the tree level
scattering amplitude derived from the lowest order effective action;
and  a correction term $-2\zeta (3)$, which gives the leading
correction to the effective action; from the kinematical factors
$K,\tilde{K}$ one finds that it corresponds to a term in the effective
action quartic in the Riemann tensor~\cite{GW} (remember that for type II
superstrings the coefficient $k$ in eq.~(\ref{act0}) is zero, 
i.e. there is no $\sim R_{\mu\nu\rho\sigma}^2$ correction).

Writing the Riemann zeta function as $\zeta
(3)=\sum_{n=1}^{\infty}n^{-3}$, we see that the correction can be
interpreted as a sum over an infinite tower of intermediate states,
and that states at level $n$ give a contribution $\sim 1/n^3$.

This means that the $\alpha '$ corrections are  dominated by
the first few massive string states. For instance, summing over the
first 10 excited states, we get $\sum_{n=1}^{10}1/n^3\simeq
1.1975\ldots$, to be compared with the value of the Riemann zeta
function $\zeta (3)\simeq 1.2020\ldots $.

Having obtained a physical picture of the mechanism that  is
responsible for the perturbative
$\alpha '$ corrections to the effective action
(integration over the first few massive modes), we proceed in the next
section to discuss the physical mechanism that is responsible for the
regularization of the singularity.
    
\subsection{Small distances vs. large energy singularities}
It can be useful to distinguish between small distance singularities
and large energy singularities. In field theory small distances means
large energies, and the distinction is therefore  meaningless. In string
theory, however, the behavior of amplitudes in high energy scattering
suggests the existence of a generalized uncertainty principle of the
form~\cite{GUP}
\be
\Delta x\ge \frac{1}{\Delta p}+{\rm const.}G_N\Delta p\, .
\ee
(Such an uncertainty principle is also suggested by  general
properties of quantum black holes~\cite{mm}.)
Therefore in string theory
the connection between high energies and small distances
is not completely trivial and it is useful to consider the two cases
separately. 

The most basic
reason why we expect strings to regulate singularities at
small distances is that the very notion of invariant point-like event
is not meaningful in string theory. While in quantum
field theory an invariant event can be defined through the splitting
of a particle in two, no such concept exist for strings, since the
point in space-time where a string splits in two depends on the
Lorentz frame used~\cite{GSW}.

At large energies, the regulating mechanism is 
instead the exponential growth
of the density of states. The asymptotic density of states has the
form
\be\label{dens}
d(M)\sim\left(\frac{M}{M_0}\right)^{-b}e^{M/M_0}\, ,
\ee
where $M_0=c/\sqrt{\alpha '}$ and $b,c$ are numerical constants, which
depend on the string theory under consideration. An example of how the
density of states prevents  quantities with
dimension of energy from growing  arbitrarily large    is given by 
the fact that we cannot raise the temperature of a system beyond
the Hagedorn temperature, $T_{\rm HAG}=M_0$,
because otherwise the canonical partition
function diverges.

Another interesting example is given by the existence of a limiting
value of the electric field for an open bosonic string. 
In the presence of an external electric field $E$,  the
rate for charged-string pair production
is in fact~\cite{BP}
\be\label{rate}
w\sim\sum_Sq_S\sum_{k=1}^{\infty}
(-1)^k\left(\frac{|\epsilon |}{k}\right)^{(d+1)/2} 
\exp\left\{ -\frac{\pi k}{|\epsilon |}(M_S^2+\epsilon^2)\right\}\, .
\ee
The sum goes over all physical string states $S$, with mass $M_S$, and 
$q_S$ is a factor that depends on the charge of the state $S$;
in the weak field limit, $\epsilon\simeq eE+o(E^3)$, where $e$ is the
total electric charge of the string, 
and one recovers the Schwinger result. As long
as $\epsilon$ is finite, the factor $\exp (-\pi kM_S^2/|\epsilon|)$ 
ensures the convergence of the rate, even if the sum over the states
$S$, for large masses, is an integral with the exponentially growing
density of states~(\ref{dens}). However, at a critical value of the
electric field, $\epsilon$ goes to infinity and  the rate therefore
diverges.

The cosmological singularity is a large energy
singularity: the Hubble constant, and therefore the curvature or the
energy density, diverges on the lowest order solutions of the equations
of motion. We therefore expect that the regulating mechanism has to do
with the existence of an infinite set of massive states; in fact, we
expect something  similar to what happens in eq.~(\ref{rate}),
since the existence of a maximum electric field and of a maximum
gravitational field do not seem to be two fundamentally different
problems. 

So, the first conclusion is that we should not necessarily expect that
perturbative $\alpha '$ corrections regulate the singularity. As we
discussed in sect.~3.1, these corrections can be accounted for with an
accuracy better than one per cent summing over the first ten mass
levels. The regularization of the singularity, instead, must be
crucially related to the existence of an infinite tower of massive
states. Of course, this does not mean that the mechanism proposed
in~\cite{GMV} cannot regulate the singularity. It simply means that we
do not have any  {\em a priori} argument that allows us to 
confidently state that the mechanism will  operate.

The second point suggested by the above considerations is to look for
massive modes production in a gravitational field. This will be
discussed in the next section.

\renewcommand{\theequation}{4.\arabic{equation}}
\setcounter{equation}{0}
\section {Production of highly excited modes}
A time-varying gravitational field produces particles.
The general mechanism is the existence of a non-trivial Bogoliubov
transformation between the in and the out vacuum.
For instance, in De~Sitter space with Hubble constant $H$, a detector
moving on a geodetic sees a thermal bath of particles 
at a temperature
\be\label{TH}
T(H)=\frac{H}{2\pi}\, .
\ee
This suggests that in De~Sitter space
a massive string mode with mass $M$ is created
with a probability  proportional  to $\exp (-M/T(H))$.
One might wonder whether a truly stringy computation of massive modes
creation could give a different answer from the field theory
result~(\ref{TH}). However, Lawrence and Martinec~\cite{LM} 
(see also~\cite{Mar}) used string
field theory to 
compute massive modes production in a FRW space with 
\be\label{Ceta}
ds^2=C(\eta )\left( d\eta^2-d{\vec{x}}^2\right)\, ,\hspace{10mm}
C(\eta )=A+B\tanh (\rho\eta )\, ,
\ee
and they obtained the same result that holds
in this model in the field theory limit, discussed e.g. in
ref.~\cite{BD}.

The above result suggests that 
the rate of energy density production due to massive modes creation
during the expansion is approximately
\bees
\dot{\rho}&\simeq &\frac{1}{\lambda_s^d}
\int_{M_0}^{\infty}dM\, M d(M) e^{-2\pi M/H}\sim\nonumber\\
\label{drho}&\sim & \frac{M_0^2}{\lambda_s^d}
\int_1^{\infty} dx\, x^{1-b}e^{-\gamma x}\, ,
\ees
where
\be\label{gam}
\gamma =\frac{2\pi M_0}{H}-1=\frac{2\pi c}{H\sqrt{\alpha '}}-1
\ee
and $d(M)$ is the density of states, given asymptotically by
eq.~(\ref{dens}). 
This formula, which has been suggested in~\cite{LM} as valid for
all the mass levels,
is also qualitatively confirmed by the explicit computation of the
Bogoliubov coefficients for the graviton production, i.e. at mass level
zero~\cite{GV,BGGV}. In this case, in fact, one finds that $|\beta_k|^2$
is a number of order 1 for energies of the order of the string mass
 (red-shifthing the frequency at the present epoch, this means
that $|\beta_k|^2=O(1)$ if the physical frequency, as seen today, is
of order 10 GHz), and has an exponential cutoff for larger
frequencies. Actually, the spectrum also shows some deviations from an
exact black body spectrum (it goes like $\omega^3$ for low frequencies
but as $\omega^{3-2\mu}$, where $\mu$ depends on details of the string
phase, for intermediate frequencies.) These deviations, however, are
not very relevant to the present discussion.\footnote{In an expanding
Universe one should also add to $\dot{\rho}$ a term $-aH\rho$
representing the dilution of string density due to the
expansion~\cite{Tur}, and $a=d-1$ for long strings.  However,
this term is not very important for the qualitative discussion
presented below, as we checked also numerically.}

Equation~(\ref{drho}) clearly has a regularizing effect on the growth of
$H$. When $H$ exceeds a critical value $H_c$, given by
$H_c/(2\pi )=M_0$, the
production rate diverges. This result can also be interpreted by noting
that $H/(2\pi )$ is the Hawking temperature in  De~Sitter
space, and therefore the condition on $H_c$ is the requirement that the
Hawking temperature does not exceed the Hagedorn temperature, 
$T_{\rm  HAG}=M_0$.  

The effect that we are discussing is non-perturbative in $\alpha
'$. In fact, for $\gamma >0$ the integral in eq.~(\ref{drho}) is an
incomplete gamma function, and from its known asymptotic expansion 
we find that, in the limit $\alpha '\ra 0$, 
\be\label{limit}
\dot{\rho}\sim\left(\frac{1}{\alpha '}\right)^{\frac{d}{2} +1}
\exp\left\{ -
\frac{2\pi c}{H\sqrt{\alpha '}}\right\}\, ,
\ee
and it is therefore non-analytic in $\alpha '$ at $\alpha '= 0$.

To understand in more detail the evolution of the cosmological model,
we  tentatively  use the expression~(\ref{drho}) for the energy
density produced, as a backreaction term in the cosmological
equations. 
The equations of motion, without perturbative $\alpha '$ corrections,
and including  matter sources with $T_{\mu}^{\nu}=(\rho
(t),-\delta_i^j p(t))$, are~\cite{GV,TV} (in the isotropic case, in $d$
spatial dimensions)
\be\label{vin2}
\dfbar^2-d H^2=2\lambda_s^{d-1}e^{\fbar}\bar{\rho}
\ee
\be
\label{dyn21}2\ddot{\fbar}-\dfbar^2-d H^2=0
\ee
\be
\label{dyn22}\dot{H}-H\dfbar =\lambda_s^{d-1}e^{\fbar}\bar{p}
\ee
where $\fbar =\phi -d\log a, e^{\fbar}\bar{\rho}=e^{\phi}\rho,
e^{\fbar}\bar{p}=e^{\phi}p$. We use  eq.~(\ref{drho})
for the time derivative of 
$\rho$; the pressure $p$ is determined by the
conservation of the energy momentum tensor, which gives
\be\label{cons}
\dot{\rho}+d H (\rho +p)=0\, . 
\ee
This is equivalent to requiring that the
constraint equation~(\ref{vin2}) is conserved by the dynamical
equations of motion~(\ref{dyn21}) and (\ref{dyn22}). 
Since we know $\dot{\rho}$ rather than $\rho ,p$, these are in principle
integro-differential equations. However,
we can combine
eq.~(\ref{dyn22}) and eq.~(\ref{vin2}) so that only the combination 
$\rho +p$ appears, which is expressed through $\dot{\rho}$ using
eq.~(\ref{cons}), and we are left with ordinary differential equations. 
Writing everything in terms of $\phi$ rather than
$\fbar$, the dynamical equations of motion can then be written as
\be\label{dyn3}
2\dot{H}+\df^2-2(d+1)H\df +d(d+1)H^2=-\frac{2}{d}\lambda_s^{d-1}e^{\phi}
\frac{\dot{\rho}(H)}{H}\, ,
\ee
\be
2\ddot{\phi}-2d\dot{H}-\df^2+2dH\df -d(d+1)H^2=0\, .
\ee
The constraint equation
\be
\df^2+d(d-1)H^2-2 d H\df =2\lambda_s^{d-1}e^{\phi}\rho
\ee
is still an integro-differential equation. However, we impose the
constraint at the initial time $t_0\ll -1$ (in units $\lambda_s=1$), 
when the right-hand side is
negligible. The constraint is then automatically preserved by the
evolution (we use its conservation as a check of the numerical
integration).  We have then
integrated  these equations numerically. The results are qualitatively
the same for any $d$ and any value of $b,c$ in 
eqs.~(\ref{drho}) and (\ref{gam}) that
we have tried.  The plots
 that we present refer to $d=3$ and to the
values $b=10, c=10$ (this value of $c$ is responsible for the  values 
$\sim O(10)$ of the $H$-scale in Fig.~1). Similar results holds for 
$c=[ (2+\sqrt{2}) \pi ]^{-1}\simeq 0.09$, which is the value 
for the heterotic string.

Figure~1 shows the behavior of the Hubble parameter versus cosmic time
$t$. At large negative values of $t$ we recover of course the lowest
order solution, since the production of massive string states is
exponentially small; $H(t)$ then reaches a maximum value and then very
abruptly bounces off and goes to zero at a finite value of $t$. At
this point the  numerical 
integration stops.\footnote{The integration stops because the integral
in eq.~(\ref{drho}) is only defined for $\gamma >0$. For negative
values of $H$ we could replace $H$ is eq.~(\ref{gam}) with
$|H|$. This is motivated by the fact that, for instance,
in the model defined in
eq.~(\ref{Ceta}) the production rate is the same independently of
whether we consider a model with a scale factor increasing in time or
decreasing in time with the same rate. However, as we will now
discuss, a more crucial modification of our equations is needed.}
The behavior of $\df$ is shown in fig.~2,
while fig.~3 shows $\dot{\rho} (t)$. 

These plots show that some important aspects of the regularization
mechanism are still missing from our equations. In fact, even if $H$
does not diverge, $\dot{H}$ becomes very large
when $H$ goes to zero, and the Ricci scalar ${\cal R}=-6(\dot{H}+
2H^2)$ in this regime is dominated by $\dot{H}$ and is large. 
The missing ingredient can be traced back to eq.~(\ref{TH}). This form
of the temperature holds for  De~Sitter space, where $H$ is constant
and $\dot{H}=0$. It is inappropriate at a stage of the cosmological
evolution when $H$ is small compared to
the critical value but $\dot{H}$ is
large. Indeed, at this stage our equations predict no massive mode
creation, while a large $\dot{H}$, as any strong time-varying
gravitational field, will produce particles, much the same as a large
$H$. This is quite clear physically, but it is anyway
interesting to check this point in a toy model. One can consider for
instance a model with a scale factor such that $H=A/\cosh
(\omega t)$, with $A,\omega >0$. 
If $A\ll 1$ but $A\omega\gg 1$ the Hubble constant is
always small during the evolution while $\dot{H}$ reaches a maximum
value $\dot{H}_{\rm max}=
A\omega/\sqrt{2}\gg 1$. A straightforward computation gives a
production rate of massive particles $\sim\exp (-M/T)$ with
$T\simeq \dot{H}_{\rm max}$.

So, at least at the level of phenomenological description of the
string phase, one should generalize eq.~(\ref{TH}) to 
\be\label{TH2}
T=T(H,\dot{H},\ddot{H},\ldots,\dot{\phi},\ddot{\phi},\ldots )\, ,
\ee
where all higher order derivatives of $H$  appear. Note that we
also included a dependence on $\dot{\phi}$ because, changing frame,
say from the string to the Einstein frame, the derivative of the dilaton
field contributes to the gravitational field, and therefore in a
generic frame even $\dot{\phi}$ must contribute to massive modes
production. For physical reasons, we expect that $T$ will be a sum
of positive definite contributions. For instance, a possible plausible
form could be
\be\label{TH3}
T=\frac{1}{2\pi}\left( H^2+c_1\dot{H}^2+c_2\dot{\phi}^2+\ldots
\right)^{1/2}\, ,
\ee
with $c_1,c_2>0$.
Now it becomes technically impossible to solve the differential
equations, since already including a term $\dot{H}$ into $T$, the
highest derivative of $H$ in the differential equations appears as
argument of the incomplete gamma functions, which makes the equations
intractable, even numerically. However, the qualitative effects of
eq.~(\ref{dyn3}), combined with an expression of the form of
eq.~(\ref{TH2},\ref{TH3}) 
for the temperature,  can be understood as follows. 

When any of the derivative of $H$  or $\phi$ exceeds a critical
value of order one in units $\lambda_s=1$, 
the right-hand side of eq.~(\ref{dyn3})
diverges, while the left-hand side is finite. So there can be no
solution where $H,\dot{H},\dot{\phi},\ldots$ are larger than their
critical value. Similarly to what happens to $H$ in fig.~1, when 
$\dot{H},\dot{\phi},$ etc. approach their respective
critical values, the effect
of the term $\dot{\rho}$ in eq.~(\ref{dyn3}) is to repel them away.
Furthermore, $H$ cannot go to zero. Indeed, unless all
higher derivatives $\dot{H},\dot{\phi},\ldots$ vanish (as it happens
at $t\ra -\infty$ if the evolution starts from the string perturbative
vacuum), then $\dot{\rho}(H,\dot{H},\dot{\phi},\ldots )$ 
does not vanish for $H\ra 0$, contrarily
to what happens when we use the expression $T(H)=H/2\pi$, see
eq.~(\ref{limit}). Therefore the right-hand side of eq.~(\ref{dyn3})
diverges if $H\ra 0$ while the left-hand side is finite, because 
$\dot{H},\dot{\phi}$ are bounded. Therefore, there can be no solution
with $H$ approaching zero. 
So, we expect a solution in which the scale factor in the string frame
is always expanding, $H>0$, possibly in a very complicated way, but
still without divergences in the curvature or in any other
quantity.\footnote{We stress that this is not a proof that there will
be a non-singular solution. In principle it may happen that at some
value of $t$ two different solutions `collide' and move into the
complex plane, as we have seen in sect.~2.3 for the equation with
perturbative $\alpha '$ corrections.}

The results obtained trying to incorporate  massive mode
production as a backreaction on the metric and dilaton field show
that massive modes may have a very important effect on the
evolution of the cosmological model.
However, the very complexity of the equations that describe the
backreaction of massive modes on the massless modes $\phi, g_{\mu\nu}$
indicates that a treatement as backreaction is not 
appropriate. If we enter
a full stringy regime, with an infinite tower of excited
massive fields,  it is not surprising that a 
description of the cosmological model in terms of classical evolution
of  massless fields runs into difficulties.

Finally we stress that, even if our results allow to obtain some
understanding of the physical mechanism that regularizes the
singularity of the lowest order solution, still this mechanism
cannot be the full
story, since it does not provide a solution that, besides being possibly 
non-singular, is also smoothly connected with the present radiation
dominated phase. This is the so-called graceful exit
problem~\cite{gra1,gra2,BM,BM2}. It can be posed as follows. From
eq.~(\ref{vin2}) one sees that there are two solutions for
$\dot{\bar{\phi}}$, given by
\be\label{branch}
\dot{\bar{\phi}}=\pm\left(
dH^2+2\lambda_s^{d-1}e^{\phi}\rho\right)^{1/2}\, .
\ee
These two solutions are usually referred to as the $(+)$ and $(-)$
branches, depending on the choice of sign in front of the square
root. The pre-big-bang solution starts on the $(+)$ branch; this
ensures that the solution start from a regime of low curvature and
couplings, and  goes
through a phase of inflation driven by the kinetic energy of the
dilaton field. The final state, however, should correspond to a
radiation dominated FRW model with constant dilaton, and therefore 
$\dot{\bar{\phi}}=\dot{\phi}-dH<0$, and we are on the $(-)$
branch. Therefore at least one branch change must take
place. Eq.~(\ref{branch}) shows that a necessary condition for this to
happen is that, at some point in the evolution, the matter density
$\rho$, in the string frame, be negative. Further global conditions
are necessary to ensure that no further branch change back to the
$(+)$ branch takes place, and have been discussed in
refs.~\cite{BM,BM2}. However, we see that, if we include massive
modes as the only source of energy density, $\rho$ is necessarily
positive and no branch change can occur. Other source of corrections,
like $\alpha '$ correction, provide an effective contribution to
$\rho$ which is negative and therefore allows branch
changes. Still, to satisfy the global conditions for a succesful exit
completion, $e^{\phi}$ corrections must also be included. An
example in which the combined effect of $\alpha '$ corrections and
$e^{\phi}$ corrections provides a successful exit has been discussed
in~\cite{BM2}. So, it is clear that the effect that we have studied in
this paper is only an ingredient 
of the mechanism that finally can produce a
well-motivated and non-singular cosmological model, that eventually
matches standard cosmology.

\renewcommand{\theequation}{5.\arabic{equation}}
\setcounter{equation}{0}
\section {Conclusions}
The lowest-order cosmological solutions obtained from the string
effective action are singular, and therefore corrections to the lowest
order action must play  a crucial role.
In string theory there are two kind of corrections, `stringy
corrections', parametrized by $\alpha '$, and loop corrections,
parametrized by  $e^{\phi}$. In this paper we have drawn attention to
the fact that $\alpha '$ corrections come into two types, perturbative
and non-perturbative, and that they have a distinct physical
origin. Perturbative corrections are related to the integration over
the first few massive modes while non-perturbative corrections comes
from the exponentially growing density of massive states.  As we
discussed in sect.~3, this
physical interpretation makes clear that non-perturbative $\alpha '$
corrections are expected to have a regularizing effect.

The explicit verification of the regularization mechanism meets
difficulties both for pertubative and non-perturbative 
$\alpha '$ corrections. In the former case, a truncation at any finite
order in perturbation theory gives scheme-dependent results, so that
an all-order computation is in principle required; 
since an all-order perturbative computation is anyway faced with the 
problem of the convergence and resummation
of the asymptotic series, this is already a
hint of the fact that the full answer requires the inclusion
of non-perturbative effects. 

In the case of non-perturbative corrections related to massive modes
production, the technical difficulty that we have
to face is the inclusion of the effect of all higher order derivative
of $H$ and $\phi$ in the computation of the production rate of massive
modes. However, independently of the technical difficulties,
the mechanism has a clear physical interpretation, which
suggests that it  regularizes the lowest order solution. 

We conclude by pointing out briefly a possible phenomenological
consequence of the  scenario in which in the string phase we have
highly excited string modes. 
The massive string states produced will eventually decay.
The decay of the excited string levels has been studied  in
ref.~\cite{WTM}. The result is that, for large level number $N$, 
the  width is dominated  by
decays in which one of the products is massless, i.e. by transitions
of the form $N\ra N-1$ with emission of a graviton or another massless
particle.  From the mass formula $M_N^2\sim N$,
the level spacing is $\Delta M_N\sim 1/M_N$. If a string
at an excited level $N$ decays at level  $N-1$
with emission of a graviton of frequency $\omega=\Delta M_N$, we get
transitions at the frequencies $f_N=f_0/\sqrt{N}$. The
most naive estimate of the frequency $f_0$, as seen today, 
is obtained by red-shifting
the string mass from a string time to the present epoch, which
gives  $f_0=\omega_{\rm today}/(2\pi )$
 of the order of $10-10^2$ GHz, which is also the typical cutoff
value of the gravitons produced by quantum vacuum
fluctuations~\cite{GV,BGGV,AB}. 
However, this radiation is  produced
within the string phase, rather than at the transition between the
string phase and the ${\em FRW}$ phase,
and a further red-shift must be included, to
take into account the expansion from the time of production to the
onset of a standard {\em FRW} cosmology. We
cannot compute it, in the absence of a knowledge of the mechanism that
terminates the string phase. There is, however, the interesting
possibility that, because of this further redshift and/or the factor
$1/\sqrt{N}$ in $f_N$,
 this gravitational radiation might be shifted  in
the frequency band accessible to the LIGO/Virgo gravitational
wave detectors, which are expected to operate in the 6~Hz--1~kHz region.

Concerning the intensity of this gravitational 
radiation, it can be estimated as follows. 
If there is not a subsequenty inflationary phase below the string
phase, also the photons that we observe today in the CMBR have  been
produced during the large curvature phase of the model, by 
massive modes decay. Since at this scale the rates for photon and
graviton production are comparable, this suggests that the overall
intensity of the relic gravitational wave background produced by the
decay of massive string modes has an energy density comparable to the
energy density of the 2.7~K radiation.

\vskip 2 cm

\section*{Acknowledgements}
I am grateful to Alessandra Buonanno, Stefano Foffa,
Krzysztof Meissner, Riccardo Sturani,
Carlo Ungarelli and Gabriele Veneziano for very useful discussions. I
am especially grateful to  Jorge Russo for invaluable discussions.
I also thank the referee for stimulating comments.

\newpage
\begin{figure}[t]
\epsfxsize=3in
\centerline{\psfig{figure=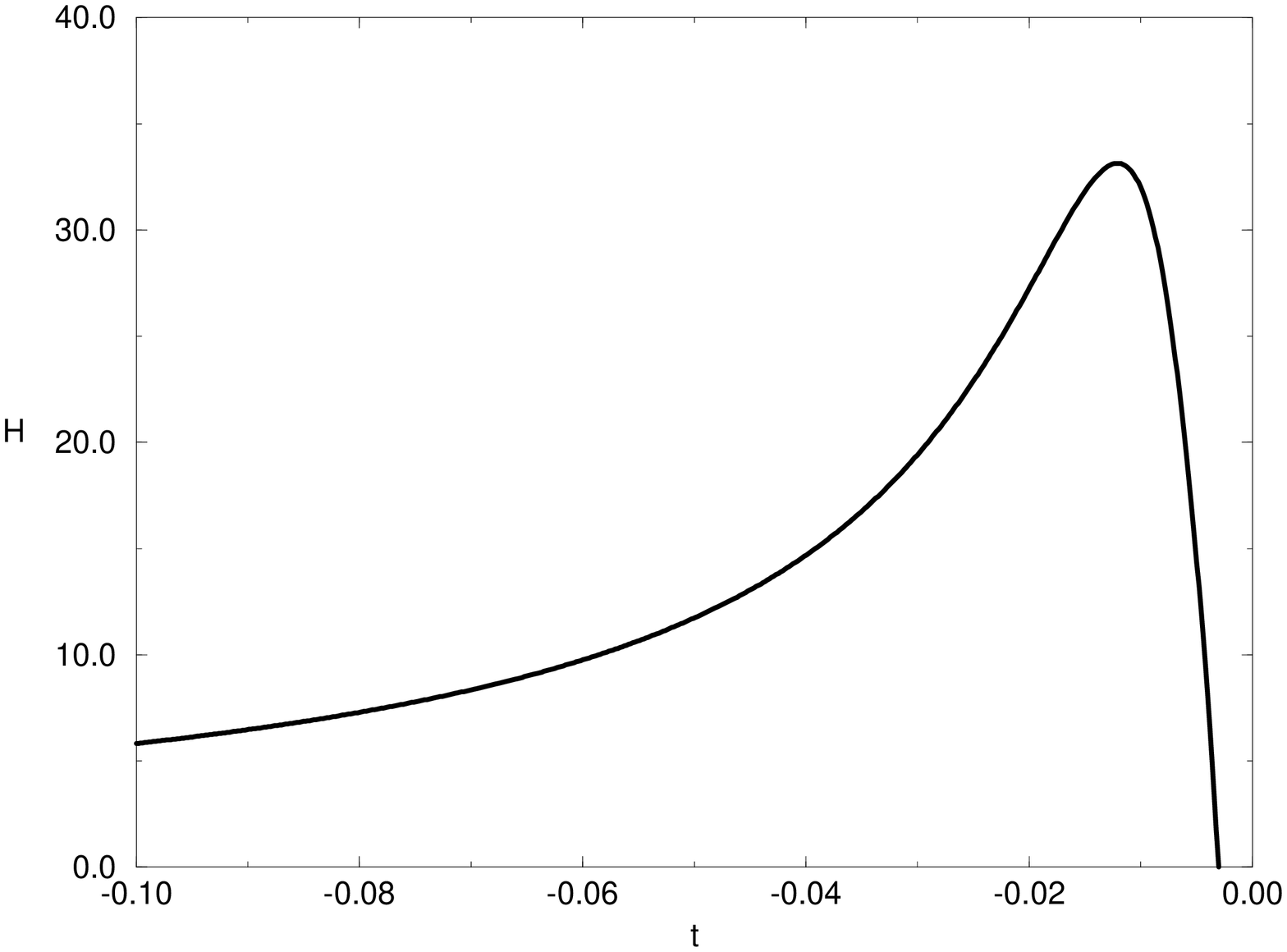,width=3in,angle=-90}}
\caption{\sl The Hubble parameter $H$ as a function of time. 
The initial conditions
for the numerical integration are  $t_0=-20,\phi (t_0)=-10$,
$H(t_0)=-1/(t_0\sqrt{3} )$; $\df (t_0)$ is determined by the constraint
equation.
} 
\end{figure}

\newpage
\begin{figure}[t]
\epsfxsize=3in
\centerline{\psfig{figure=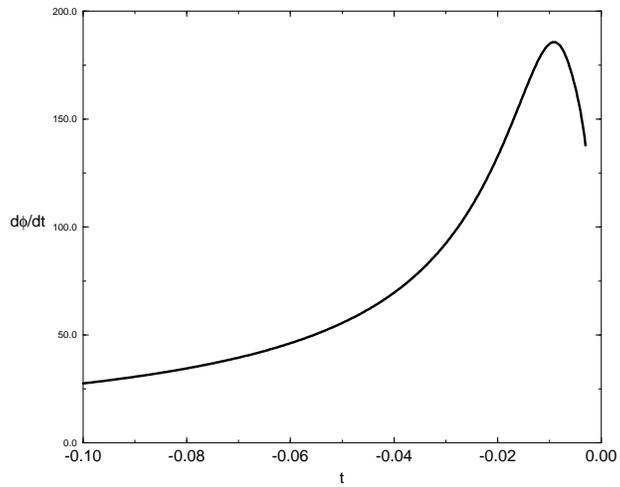,width=3in,angle=-90}}
\caption{\sl  $\df$ as a function of time. Same initial conditions as in
  fig.~1. The integration stops when $H$ reaches zero.} 
\end{figure}

\newpage
\begin{figure}[t]
\epsfxsize=3in
\centerline{\psfig{figure=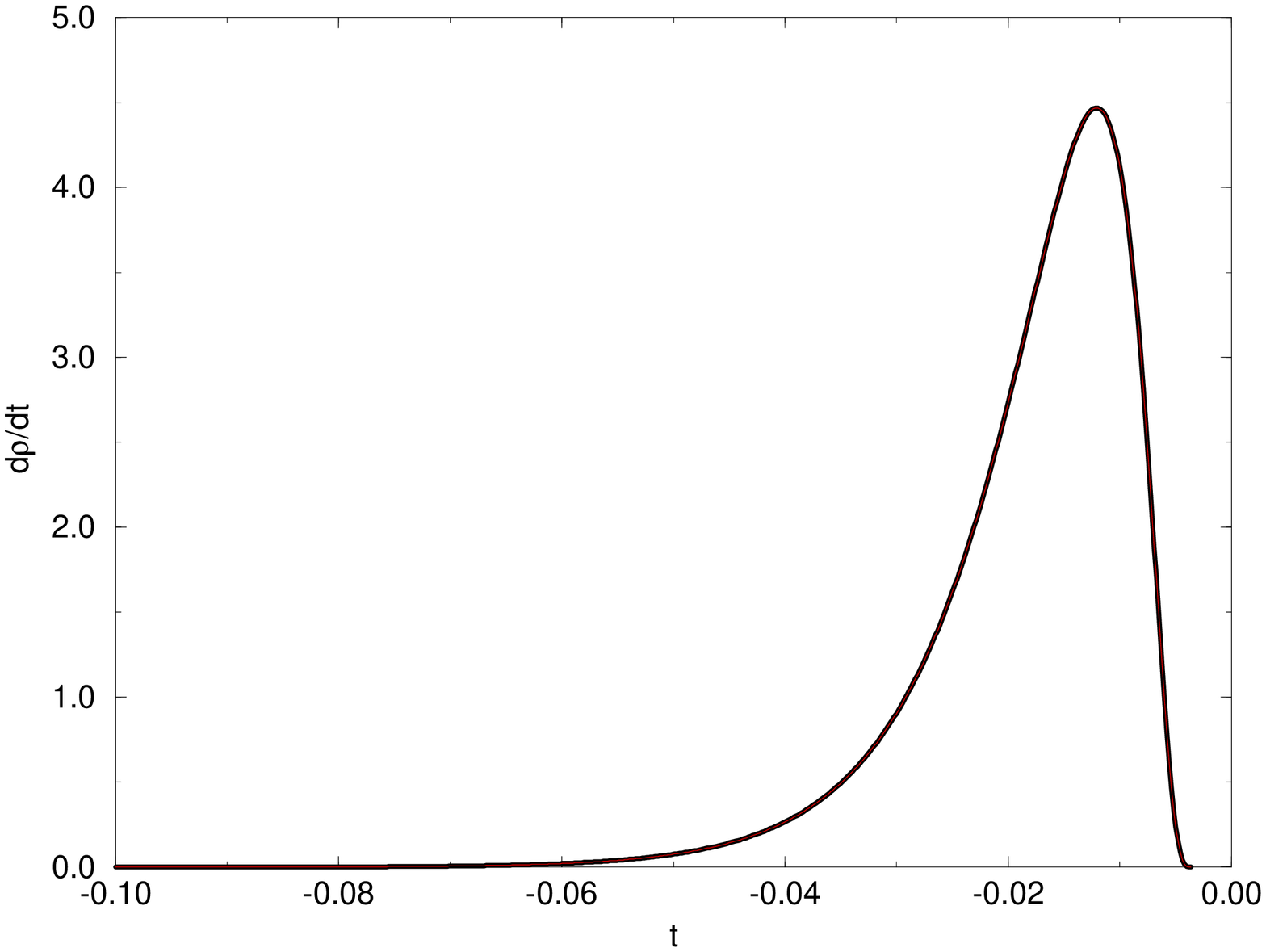,width=3in,angle=-90}}
\caption{\sl  $\dot{\rho}$ as a function of time. Same initial conditions as in
  fig.~1.} 
\end{figure}


\newpage

\begin{thebibliography}{99}
\bibitem{Ven} G. Veneziano, Phys. Lett. B265 (1991) 287.
\bibitem{GV} M. Gasperini and G. Veneziano, Astropart. Phys. 1 (1993)
317; Mod. Phys. Lett. A8 (1993) 3701; Phys. Rev. D50 (1994) 2519.
\bibitem{review} G. Veneziano,
in {\em ``String Gravity and Physics at the Planck
Energy Scale''}, N.~Sanchez and A.~Zichichi eds.,
Kluwer Publ., Dordrecht, 1996, pg. 285; M. Gasperini, ibid. pg. 305.

An up-to-date collection of references on string cosmology can be found
at http://www.to.infn.it/teorici/gasperini/
\bibitem{sc} R. Brandenberger and C. Vafa, Nucl. Phys. B316 (1989) 391;

B. Campbell, A. Linde and K. Olive, Nucl. Phys. B355 (1991) 146;

E. Copeland, A. Lahiri and D. Wands, Phys. Rev. D50 (1994) 4868;

I. Antoniadis, J. Rizos and K.~Tamvakis, Nucl. Phys. B415 (1994) 497.

\bibitem{TV} A. Tseytlin and C. Vafa, Nucl. Phys. B372 (1992) 443.

\bibitem{duality} K.A. Meissner and G. Veneziano, Phys. Lett. B267 (1991) 33;
Mod. Phys. Lett. A6 (1991) 3397;

M. Gasperini and G. Veneziano, Phys. Lett. B277 (1992) 256;

A. Tseytlin, Mod. Phys. Lett. A6 (1991) 1721;

A. Sen, Phys. Lett. B271 (1991) 295.
\bibitem{GMV} M. Gasperini, M.~Maggiore and G.~Veneziano,
  Nucl. Phys. B494 (1997) 315.
\bibitem{gra1} R. Brustein and G. Veneziano, Phys. Lett. B329
(1994) 429; N. Kaloper, R. Madden and K.~Olive, Nucl. Phys. B452
(1995) 677.
\bibitem{gra2}  M. Gasperini, J. Maharana and G. Veneziano,
Nucl. Phys. B472 (1996) 349;

S. Rey, Phys. Rev. Lett. 77 (1996) 1929;  hep-th/9609115;

M. Gasperini and G. Veneziano, Phys. Lett. B387 (1996) 715;

A.~Buonanno, M.~Gasperini, M.~Maggiore and C.~Ungarelli,
Class. Quant. Grav. 14 (1997) L97;

\bibitem{BM} R. Brustein and R. Madden, Phys. Lett. B410 (1997) 110.
\bibitem{BM2} R. Brustein and R. Madden, Phys. Rev. D57 (1998) 712.
\bibitem{azione} C. Callan, D. Friedan, E.~Martinec and
  M.~Perry, Nucl. Phys. B262 (1985) 593.
\bibitem{FT} E. Fradkin and A. Tseytlin, Phys. Lett. B158 (1985) 316;
Nucl. Phys. B261 (1985) 1.
\bibitem{GW} D. Gross and E. Witten, Nucl. Phys. B277 (1986) 1.
\bibitem{Zwi} B. Zwiebach, Phys. Lett. B156 (1985) 315.
\bibitem{MT} R. Metsaev and A. Tseytlin, Nucl. Phys. B293 (1987) 385.
\bibitem{DR} S. Deser and A. Redlich, Phys. Lett. B176 (1986) 350;
\bibitem{FOTW} K.~F\"orger, B. Ovrut, S.~Theisen and D.~Waldram, Phys. Lett. B388
(1996) 512.
\bibitem{Mei} K. Meissner, Phys. Lett. B392 (1997) 298.
\bibitem{MK} N. Kaloper and K.~Meissner, 
Phys. Rev. D56 (1997) 7940.
\bibitem{BNY} R. Brustein, D. Nemeschansky and S. Yankielowicz,
Nucl. Phys. B301 (1988) 224.
\bibitem{Tse} A.A. Tseytlin, Phys. Lett. B185 (1987) 59.
\bibitem{BFLP} I.L.~Buchbinder, E.S.~Fradkin, S.L.~Lyakhovich and
V.D.~Pershin, Phys. Lett. B304 (1993) 239.
\bibitem{GS} M. Green and J. Schwarz, Nucl. Phys. B181 (1981) 502,
  B198 (1982) 441.
\bibitem{GUP} G. Veneziano, Europhys. Lett. 2 (1986) 199;

D. Amati, M. Ciafaloni and G.~Veneziano, Phys. Lett  B216 (1989) 41,
B197 (1987) 81;
Int.~J.~Mod. Phys. A3 (1988) 1615;  Nucl. Phys. B347 (1990) 530;

D.J. Gross and P.F. Mende, Phys. Lett.  B197 (1987) 129; 
Nucl. Phys. B303 (1988) 407;

K. Konishi, G. Paffuti and P.~Provero, Phys. Lett. B234 (1990) 276.
\bibitem{mm} M. Maggiore, Phys. Lett. B304 (1993) 65.
\bibitem{GSW} M. Green, J. Schwarz and E. Witten, {\em Superstring
Theory}, Cambridge Univ. Press, Cambridge, 1987, Vol.~1, pg.~29.
\bibitem{BP} C. Bachas and M. Porrati, Phys. Lett. B296 (1992) 77.
\bibitem{LM} A. Lawrence and E. Martinec, Class. Quantum Grav. 13
  (1996) 63.
\bibitem{Mar} E. Martinec, Class. Quant. Grav. 12 (1995) 941.
\bibitem{BD} N. Birrel and P.C.W. Davies, {\em Quantum fields in
curved space}, Cambridge University Press, Cambridge 1982.
\bibitem{BGGV} R. Brustein, M. Gasperini, M. Giovannini and
  G.~Veneziano,  Phys. Lett B361 (1995) 45;

R. Brustein, M. Gasperini and G. Veneziano, Phys. Rev. D55 (1997)
3882;

A. Buonanno, M. Maggiore and C. Ungarelli,   Phys. Rev. D55 (1997)
3330;
\bibitem{Tur} N. Turok, Phys. Rev. Lett. 60 (1988) 549.
\bibitem{WTM} R.B. Wilkinson, N. Turok and D. Mitchell,
Nucl. Phys. B332 (1990) 131.
\bibitem{AB} B. Allen and R. Brustein, Phys. Rev. D55 (1997) 3260.
\end{thebibliography}
\end{document}